\documentclass{llncs}
\usepackage[utf8]{inputenc}
\usepackage{epsfig}
\usepackage{multicol}
\usepackage{multirow}
\usepackage{wrapfig}
\usepackage{fancyvrb}
\usepackage{url}
\usepackage{xcolor}
\usepackage{nicefrac}
\definecolor{light-gray}{gray}{0.85}
\newcommand{\bg}[1]{\colorbox{light-gray}{#1}}
\newcommand{\bw}[1]{\colorbox{white}{#1}}
\newcommand{\verbatimproperties}{\renewcommand{\baselinestretch}{0.85} \small}

\begin{document}

\title{On Comparing Alternative Splitting Strategies for Or-Parallel Prolog Execution on Multicores}

\author{Rui Vieira \and Ricardo Rocha \and Fernando Silva}

\institute{CRACS \& INESC TEC, Faculty of Sciences, University of Porto\\
           Rua do Campo Alegre, 1021/1055, 4169-007 Porto, Portugal\\
           \email{\{revs,ricroc,fds\}@dcc.fc.up.pt}}

\maketitle


\begin{abstract}
  Many or-parallel Prolog models exploiting implicit parallelism have
  been proposed in the past. Arguably, one of the most successful
  models is \emph{environment copying} for shared memory
  architectures. With the increasing availability and popularity of
  multicore architectures, it makes sense to recover the body of
  knowledge there is in this area and re-engineer prior computational
  models to evaluate their performance on newer architectures. In this
  work, we focus on the implementation of splitting strategies for
  or-parallel Prolog execution on multicores and, for that, we develop
  a framework, on top of the YapOr system, that integrates and
  supports five alternative splitting strategies. Our implementation
  shares the underlying execution environment and most of the data
  structures used to implement or-parallelism in YapOr. In particular,
  we took advantage of YapOr's infrastructure for incremental copying
  and scheduling support, which we used with minimal modifications. We
  thus argue that all these common support features allow us to make a
  first and fair comparison between these five alternative splitting
  strategies and, therefore, better understand their advantages and
  weaknesses.
\end{abstract}


\section{Introduction}

Detecting parallelism is far from a simple task, specially in the
presence of irregular parallelism, but it is commonly left to
programmers. Research effort has been made towards making specialized
run-time systems more capable of transparently exploring available
parallelism, thus freeing programmers from such cumbersome
details. Prolog programs naturally exhibit \emph{implicit parallelism}
and are thus highly amenable for automatic exploitation. 

One of the most noticeable sources of parallelism in Prolog programs
is called \emph{or-parallelism}. Or-parallelism arises from the
simultaneous evaluation of a subgoal call against the clauses that
match that call. When implementing or-parallelism, a main difficulty
is how to efficiently represent the \emph{multiple bindings} for the
same variable produced by the parallel execution of alternative
matching clauses. One of the most successful models is
\emph{environment copying}~\cite{Ali-90a,Rocha-99b}, that has been
efficiently used in the implementation of or-parallel Prolog systems
on shared memory architectures. Recent advances in computer
architectures have made our personal computers parallel with multiple
cores sharing the main memory. Multicores and clusters of multicores
are now the norm and, although, many parallel Prolog systems have been
developed in the past, evaluating their performance or even the
implementation of newer computational models specialized for the
multicores is still open to further research.

Another major difficulty in the implementation of any parallel system
is to design efficient \emph{scheduling strategies} to assign
computing tasks to workers waiting for work. A parallel Prolog system
is no exception as the parallelism that Prolog programs exhibit is
usually highly irregular. Achieving the necessary cooperation,
synchronization and concurrent access to shared data structures among
several workers during execution is a difficult task. For environment
copying, scheduling strategies based on \emph{bottommost dispatching
  of work} have proved to be more efficient than topmost
strategies~\cite{Ali-90b}. An important mechanism that suits
bottommost strategies best is \emph{incremental
  copying}~\cite{Ali-90a}, an optimized copy mechanism that avoids
copying the whole stacks when sharing work. \emph{Stack
  splitting}~\cite{Gupta-99,Pontelli-06} is an extension to the
environment copying model that provides a simple, clean and efficient
method to accomplish work splitting among workers. It successfully
splits the computation task of one worker in two complementary sets,
and was thus first introduced aiming at distributed memory
architectures~\cite{Villaverde-01,Rocha-03a}.

In this work, we focus on the implementation of splitting strategies
for or-parallel Prolog execution on multicore architectures and, for
that, we present a framework, on top of the YapOr
system~\cite{Rocha-99b}, that integrates and supports five alternative
splitting strategies. We used YapOr's original splitting
strategy~\cite{Rocha-99b} and two splitting strategies from previous
work~\cite{Vieira-12}, named \emph{vertical} and \emph{half
  splitting}, that split work based on choice points, together with
the new implementation of two alternative stack splitting strategies,
named \emph{horizontal}~\cite{Gupta-99} and \emph{diagonal
  splitting}~\cite{Rocha-03a}, in which the split is based on the
unexplored alternative matching clauses. All implementations take full
advantage of the state-of-the-art fast and optimized Yap Prolog
engine~\cite{CostaVS-12} and share the underlying execution
environment and most of the data structures used to implement
or-parallelism in YapOr. In particular, we took advantage of YapOr's
infrastructure for incremental copying and scheduling support, which
we used with minimal modifications. We thus argue that all these
common support features allow us to make a first and fair comparison
between these five alternative splitting strategies and, therefore,
better understand their advantages and weaknesses.

The remainder of the paper is organized as follows. First, we
introduce some background about environment copying, stack splitting
and YapOr's scheduler. Next, we describe the five alternative
splitting strategies and discuss their major implementation issues in
YapOr. We then present experimental results on a set of well-known
benchmarks and advance some conclusions and further work.


\section{Environment Copying}

In the environment copying model, each worker keeps a separate copy of
its own environment, thus enabling it to freely store assignments to
shared variables without conflicts. Every time a worker shares work
with another worker, all the execution stacks are copied to ensure
that the requesting worker has the same environment state down to the
search tree node\footnote{At the engine level, a search tree node
  corresponds to a choice point in the stack.} where the sharing
occurs. To reduce the overhead of stack copying, an optimized copy
mechanism called \emph{incremental copy}~\cite{Ali-90a} takes
advantage of the fact that the requesting worker may already have
traversed one part of the path being shared. Therefore, it does not need
to copy the stacks referring to the whole path from root, but only the
stacks starting from the youngest node common to both workers.

As a result of environment copying, each worker can proceed with the
execution exactly as a sequential engine, with just minimal
synchronization with other workers. Synchronization is mostly needed
when updating scheduling information and when accessing shared nodes
in order to ensure that unexplored alternatives are only exploited by
one worker. Shared nodes are represented by \emph{or-frames}, a data
structure that workers must access, with mutual exclusion, to obtain
the unexplored alternatives. All other data structures, such as the
environment, the heap, and the trail do not require synchronization.


\section{Stack Splitting}

Stack splitting was first introduced to target distributed memory
architectures, thus aiming to reduce the mutual exclusion requirements
of environment copying when accessing shared nodes of the search
tree. It accomplishes this by defining simple, clean and efficient
work splitting strategies in which all available work is statically
divided in two \emph{complementary sets} between the sharing
workers. In practice, stack splitting is a refined version of the
environment copying model, in which the synchronization requirement
was removed by the preemptive split of all unexplored alternatives at
the moment of sharing. The splitting is such that both workers will
proceed, each executing its branch of the computation, without any
need for further synchronization when accessing shared nodes.

\begin{figure}[ht]
\centering
\includegraphics[width=12cm]{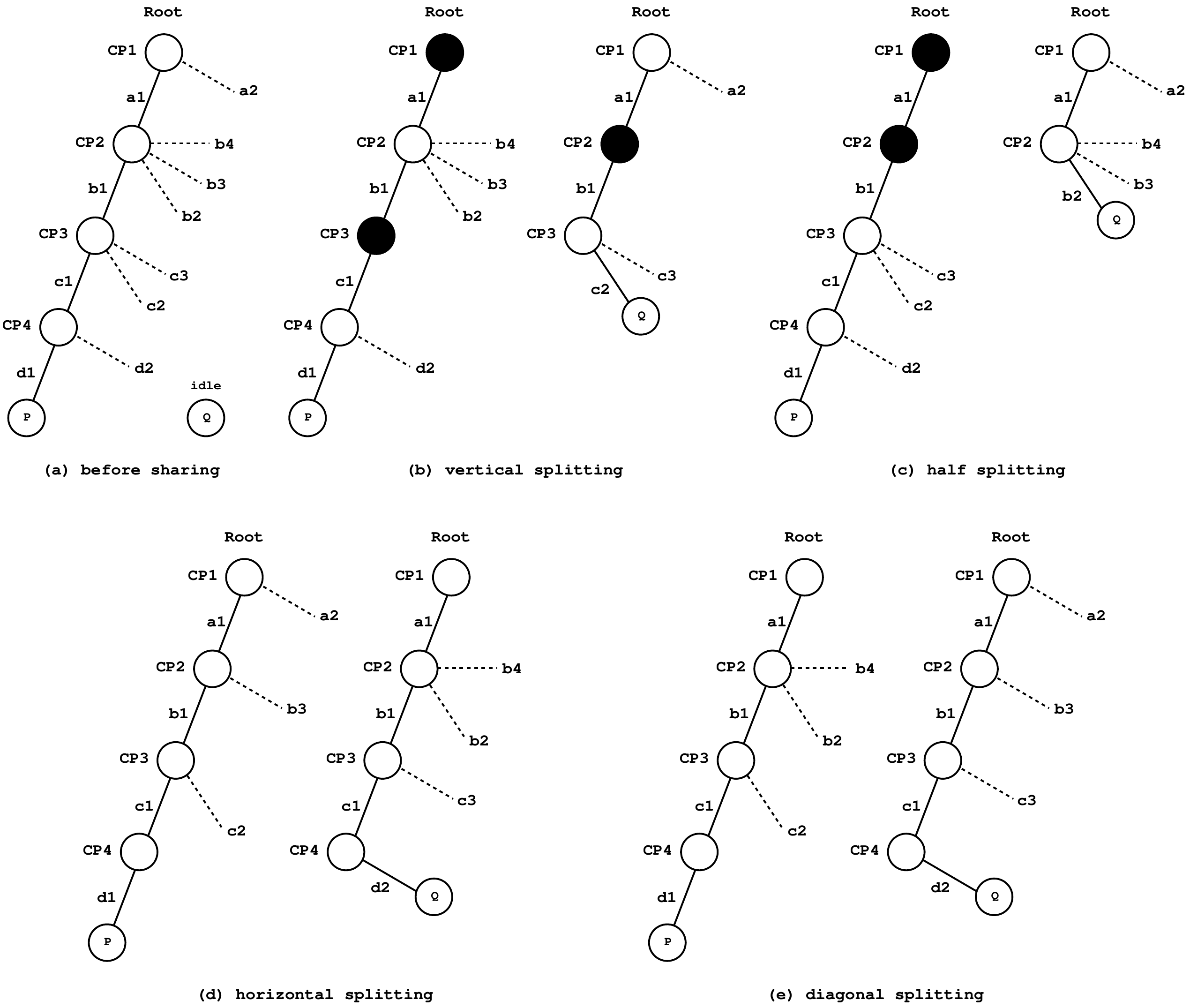}
\caption{Alternative stack splitting strategies}
\label{fig_splitting_strategies}
\end{figure}

The original stack splitting proposal~\cite{Gupta-99} introduces two
strategies for dividing work: \emph{vertical splitting}, in which the
available choice points are alternately divided between the two
sharing workers, and \emph{horizontal splitting}, which alternately
divides the unexplored alternatives in each available choice
point. \emph{Diagonal splitting}~\cite{Rocha-03a} is a more elaborated
strategy that achieves a precise partitioning of the set of unexplored
alternatives. It is a kind of mix between horizontal and vertical
splitting, where the set of all unexplored alternatives in the
available choice points is alternately divided between the two sharing
workers. Another splitting strategy~\cite{Villaverde-03}, which we
named \emph{half splitting}, splits the available choice points in two
halves. Figure~\ref{fig_splitting_strategies} illustrates the effect
of these strategies in a work sharing operation between a busy worker
$P$ and an idle worker $Q$.

Figure~\ref{fig_splitting_strategies}(a) shows the initial
configuration with the idle worker $Q$ requesting work from a busy
worker $P$ with 7 unexplored alternatives in 4 choice
points. Figure~\ref{fig_splitting_strategies}(b) shows the effect of
vertical splitting, in which $P$ keeps its current choice point and
alternately divides with $Q$ the remaining choice points up to the
root choice point. Figure~\ref{fig_splitting_strategies}(c)
illustrates the effect of half splitting, where the bottom half is for
worker $P$ and the half closest to the root is for worker
$Q$. Figure~\ref{fig_splitting_strategies}(d) details the effect of
horizontal splitting, in which the unexplored alternatives in each
choice point are alternately split between both workers, with workers
$P$ and $Q$ owning the first unexplored alternative in the even and
odd choice points,
respectively. Figure~\ref{fig_splitting_strategies}(e) describes the
diagonal splitting strategy, where the unexplored alternatives in all
choice points are alternately split between both workers in such a way
that, in the worst case, $Q$ may stay with one more alternative than
$P$. For all strategies, the corresponding execution stacks are first
copied to $Q$, next both $P$ and $Q$ perform splitting, according to
the splitting strategy at hand, and then $P$ and $Q$ are set to
continue execution. As we will see, in some situations, there is no
need for any copy at all, and a backtracking action is enough to place
the requesting worker ready for execution.


\section{YapOr's Scheduler and Original Splitting Strategy}

We can divide the execution time of a worker in two modes:
\emph{scheduling mode} and \emph{engine mode}. A worker enters in
scheduling mode whenever it runs out of work and calls the scheduler
to search for available work. As soon as it gets a new piece of work,
it enters in engine mode and runs like a sequential engine.


\subsection{Work Scheduling}

In YapOr, when a worker runs out of work, first the scheduler tries to
select a busy worker with excess of \emph{work load} to share
work. The work load is a measure of the amount of unexplored
alternatives in private nodes. There are two alternatives to search
for busy workers in the search tree: search \emph{below} or search
\emph{above} the current node where the idle worker is
positioned. Idle workers always start to search below the current
node, and only if they do not find any busy worker there, they search
above. The main advantage of selecting a busy worker below instead of
above is that the idle worker can request immediately the sharing
operation, because its current node is already common to the busy
worker, which avoids backtracking in the tree and undoing variable
bindings.

When the scheduler does not find any busy worker with excess of work
load, it tries to move the idle worker to a better position in the
search tree. By default, the idle worker backtracks until it reaches a
node where there is at least one busy worker below. Another option is
to backtrack until reaching the node that contains all the busy
workers below. The goal of these strategies is to distribute the idle
workers in such a way that the probability of finding, as soon as
possible, busy workers with excess of work below is substantially
increased.


\subsection{Work Sharing}

Similarly to the Muse system\cite{Ali-90b}, YapOr also follows a
\emph{bottommost work sharing strategy}. Whenever an idle worker
\emph{Q} makes a work request to a busy worker \emph{P}, the work
sharing operation is activated to \emph{share all private nodes} of
\emph{P} with \emph{Q}. \emph{P} accepts the work request only if its
work load is above a given \emph{threshold value}. In YapOr,
accomplishing this operation involves the following stages:

\begin{description}
\item[Sharing loop.] This stage handles the sharing of \emph{P}'s
  private nodes. For each private node, a new or-frame is allocated
  and the access to the unexplored alternatives, previously done
  through the \texttt{CP\_alt} fields in the private choice points, is
  moved to the \texttt{OrFr\_alt} fields in the new or-frames. All
  nodes have now a corresponding or-frame, which are sequentially
  chained through the fields \texttt{OrFr\_next} and
  \texttt{OrFr\_nearest\_livenode}. The
  \texttt{OrFr\_nearest\_livenode} field is used to optimize the
  search for shared work. The membership field \texttt{OrFr\_members},
  which defines the set of workers that own or act upon a node, is
  also initialized to indicate that \emph{P} and \emph{Q} are sharing
  the corresponding choice points.
\item[Membership update.] Next, the old or-frames on \emph{P}'s branch
  are updated to include the requesting worker \emph{Q} in the
  membership field (frames starting from \emph{P}'s current
  \texttt{top\_or\_frame} til \emph{Q}'s \texttt{top\_or\_frame}). In
  order to delimit the shared region of the search tree, each worker
  maintains two important variables, named \texttt{top\_cp} and
  \texttt{top\_or\_frame}, that point, respectively, to the youngest
  shared choice point and to the youngest or-frame\footnote{Please
    note that the use of the naming \emph{top} in these two variables
    can be confusing since, due to historical reasons, it refers to
    the top of the choice-point stack (where the root node is at the
    bottom) and not to the top of the search tree (where the root node
    is at the top). Despite this naming, our discussion keeps
    following a search tree approach with the root node always at the
    top.}.
\item[Compute top or-frames.] Finally, the new top or-frames in each
  worker are set, and since all shared work is available to both
  workers, both get the same \texttt{top\_or\_frame}. As we will see
  next, this is not the case for stack splitting, and the
  \texttt{top\_or\_frame} variable of \emph{Q} is set accordingly to
  the splitting strategy being considered.
\end{description}


\section{Supporting Alternative Splitting Strategies in YapOr}

Extending YapOr to support different stack splitting strategies required
some modifications to the way unexplored alternatives are accessed. In
more detail:

\begin{itemize}
\item With stack splitting, each worker has its own work chaining
  sequence. Hence, the control and access to the unexplored
  alternatives returned to the \texttt{CP\_alt} choice point fields and
  the \texttt{OrFr\_alt} and \texttt{OrFr\_nearest\_livenode} or-frame
  fields were simply ignored.
\item For the vertical and half splitting strategies, the
  \texttt{OrFr\_nearest\_livenode} field was recovered as a way to
  implement the chaining sequence of choice points. At work sharing,
  each worker adjusts its \texttt{OrFr\_nearest\_livenode} fields so
  that two separate chains are built corresponding to the intended
  split of the work.
\item In order to reuse YapOr's infrastructure for incremental copying
  and scheduling support, the or-frames are still chained through the
  \texttt{OrFr\_next} fields and still use the \texttt{OrFr\_member}
  fields for work scheduling.
\end{itemize}

Next, we detail the implementation of the vertical, half, horizontal
and diagonal splitting strategies as well as the incremental copy
technique.


\subsection{Vertical Splitting}

The vertical splitting strategy follows a pre-determined work
splitting scheme in which the chain of available choice points is
alternately divided between the two sharing workers. At the
implementation level, we use the \texttt{OrFr\_nearest\_livenode}
field in order to generate two alternated chain sequences in the
or-frames, and thus divide the available work in two independent
execution paths. Workers can share the same or-frames but they have
their own independent path without caring for the or-frames not
assigned to them. Figure~\ref{code_vertical_splitting} presents the
pseudo-code that implements the work sharing procedure for vertical
splitting.

\begin{figure}[ht]
\verbatimproperties
\begin{verbatim}
next_fr = NULL
nearest_fr = NULL
current_cp = B  // B points to the youngest choice point
while (current_cp != top_cp)  // loop until the youngest shared choice point
  current_fr = alloc_or_frame(current_cp)
  add_member(P, OrFr_member(current_fr))
  if (next_fr)
    OrFr_next(next_fr) = current_fr
    add_member(Q, OrFr_member(current_fr))
  if (nearest_fr)
    OrFr_nearest_livenode(nearest_fr) = current_fr
  nearest_fr = next_fr
  next_fr = current_fr
  current_cp = CP_b(current_cp)  // next choice point on stack
 
// connecting with the older or-frames
if (next_fr)
  if (top_or_frame == root_frame)
    OrFr_nearest_livenode(next_fr) = DEAD_END
  else
    OrFr_nearest_livenode(next_fr) = top_or_frame
  OrFr_next(next_fr) = top_or_frame
if (nearest_fr)
  if (top_or_frame == root_frame)
    OrFr_nearest_livenode(nearest_fr) = DEAD_END
  else
    OrFr_nearest_livenode(nearest_fr) = top_or_frame

// continuing vertical splitting
if (next_fr = NULL)
  current_fr = top_or_frame
nearest_fr = OrFr_nearest_livenode(current_fr)
while (nearest_fr != DEAD_END)
  OrFr_nearest_livenode(current_fr) = OrFr_nearest_livenode(nearest_fr)
  current_fr = nearest_fr
  nearest_fr = OrFr_nearest_livenode(current_fr)
\end{verbatim}
\caption{Work sharing with vertical splitting}
\label{code_vertical_splitting}
\end{figure}

The work sharing procedure starts from \emph{P}'s youngest choice
point (register \texttt{B}) and traverses all \emph{P}'s private
choice points to create a corresponding or-frame by calling the
\texttt{alloc\_or\_frame()} procedure. In
Fig.~\ref{code_vertical_splitting}, the \texttt{current\_fr},
\texttt{next\_fr} and \texttt{nearest\_fr} variables represent,
respectively, the or-frame allocated in the current step, the or-frame
allocated in the previous step, which is used to link to the current
or-frame by the \texttt{OrFr\_next} field, and the or-frame allocated
before the \texttt{next\_fr}, which is used as a double spaced frame
marker in order to initiate the \texttt{OrFr\_nearest\_livenode}
fields. For the youngest choice point, the or-frame is initialized
with just the owning worker \emph{P} in the membership field. The
other or-frames are initialized with both workers \emph{P} and
\emph{Q}.

Next, follows the connection with the older and already stored
or-frames. Here, consideration must be given to the condition of
\emph{P}'s current \texttt{top\_or\_frame}. If it is the root
or-frame, the \texttt{OrFr\_nearest\_livenode} fields of the new
or-frames are assigned to a \texttt{DEAD\_END} value, which marks the
ending point for unexplored work. Otherwise, they are assigned to
\emph{P}'s current \texttt{top\_or\_frame}.

Finally, we need to decide where to continue the vertical splitting
algorithm for the older shared nodes. If no private work was shared,
which means that we are only sharing work from the old shared nodes,
the starting or-frame is \emph{P}'s current
\texttt{top\_or\_frame}. Otherwise, if some new or-frame was created,
the starting or-frame is the last created frame in the sharing loop
stage, which was connected to \emph{P}'s current
\texttt{top\_or\_frame} in the previous step. Either way, this serves
the decision to elect the or-frame where the continuation of vertical
splitting, guided through the \texttt{OrFr\_nearest\_livenode} field,
should continue. The procedure then traverses the old shared frames
until a \texttt{DEAD\_END} is reached and, at each frame, lies a
reconnection process of the \texttt{OrFr\_nearest\_livenode} field.


\subsection{Half Splitting}

The half splitting strategy partitions the chain of available choice
points in two consecutive and almost equally sized parts, which are
chained through the \texttt{OrFr\_nearest\_livenode} field of the
corresponding or-frames. For that, the choice points are numbered
sequentially and independently per worker to allow the calculation of
the \emph{relative depth} of the worker's assigned choice points. In
order to support this numbering of nodes, a new \emph{split counter}
field, named \texttt{CP\_sc}, was introduced in the choice point
structure. Figure~\ref{code_half_splitting} presents the pseudo-code
that implements work sharing with horizontal splitting.

\begin{figure}[ht]
\verbatimproperties
\begin{verbatim}
// updating the split counter
current_cp = B  // B points to the youngest choice point
split_number = CP_sc(current_cp) / 2
while (CP_sc(current_cp) != split_number + 1)
  CP_sc(current_cp) = CP_sc(current_cp) - split_number
  current_cp = CP_b(current_cp)  // next choice point on stack
CP_sc(current_cp) = 1  // middle choice point

// assign the remaining choice points to the requesting worker
middle_fr = CP_or_fr(current_cp)
if (middle_fr)
  OrFr_nearest_livenode(middle_fr) = DEAD_END
  current_fr = top_or_frame  // top_or_frame points to the youngest or-frame
  while (current_fr != middle_fr)
    remove_member(Q, OrFr_member(current_fr))
    current_fr = OrFr_next(current_fr)
else
  // sharing loop stage
\end{verbatim}
\caption{Work sharing with half splitting}
\label{code_half_splitting}
\end{figure}

The work sharing procedure starts from \emph{P}'s youngest choice
point and updates the split counter on half of the choice points, in
decreasing order, until reaching the \emph{middle choice point} in
\emph{P}'s initial partition, which gets a split counter value of
1. These are the half choice points that, after sharing, will be still
owned by \emph{P}. The other half will be assigned to the requesting
worker \emph{Q}.

After updating the split counter, we can distinguish two different
situations. The first situation occurs when there are more old shared
choice points than private in \emph{P}'s branch, in which case the
middle choice point is already assigned with an or-frame. Thus, there
is no need for the sharing loop stage, the middle frame is assigned to
a \texttt{DEAD\_END}, to mark the end of \emph{P}'s newly assigned
work, and the requesting worker \emph{Q} is excluded from all
or-frames from the top frame til the middle frame. The second
situation occurs when the middle choice point is private, in which
case the remaining choice points are updated to belong to \emph{Q},
which includes allocating and initializing the corresponding
or-frames.


\subsection{Horizontal Splitting}

In the horizontal splitting strategy, the unexplored alternatives are
alternately divided in each choice point. For that, the choice points
include an extra field, named \texttt{CP\_offset}, that marks the
offset of the next unexplored alternative belonging to the choice
point. When allocating a private choice point, \texttt{CP\_offset} is
initialized with a value of 1, meaning that the next alternative to be
taken has a displacement of 1 in the list of unexplored
alternatives. This is the usual and expected behavior for private
choice points.

When sharing work, we follow YapOr's default splitting strategy where
a new or-frame is allocated for each private choice point in \emph{P}
and then all or-frames are updated to include the requesting worker
\emph{Q} in the membership field. Next, to implement the splitting
process, we double the value of the \texttt{CP\_offset} field in each
shared choice point, meaning that the next alternative to be taken in
the choice point is displaced two positions relatively to the previous
value. Finally, we adjust the first alternative at each choice point
for the workers \emph{P} and \emph{Q}. Recall from
Fig.~\ref{fig_splitting_strategies} that $P$ must own the first
unexplored alternative in the even choice points and $Q$ the first
unexplored alternative in the odd choice
points. Figure~\ref{code_horizontal_splitting} shows the pseudo-code
for this procedure.

\begin{figure}[ht]
\verbatimproperties
\begin{verbatim}
// the sharing worker P starts the adjustment
if (sharing worker) adjust = TRUE else adjust = FALSE
current_cp = top_cp
while(current_cp != root_cp)  // loop until the root choice point
  alt = CP_alt(current_cp)
  if (alt != NULL)
    offset = CP_offset(current_cp)   
    CP_offset(current_cp) = offset * 2
    if (adjust)
      CP_alt(current_cp) = get_next_alternative(alt, offset)
  current_cp = CP_b(current_cp)  // next choice point on stack
  adjust = !adjust
\end{verbatim}
\caption{Work sharing with horizontal splitting}
\label{code_horizontal_splitting}
\end{figure}


\subsection{Diagonal Splitting}

Diagonal splitting is an alternative strategy that implements a better
overall distribution of unexplored alternatives between
workers. Diagonal splitting is based on the alternated division of
\emph{all} alternatives, regardless of the choice points they belong
to. This strategy also follows YapOr's default splitting strategy and
uses the same offset multiplication approach as presented for
horizontal splitting, but takes into account the number of unexplored
alternatives in a choice point to decide how the partitioning will be
done in the next choice point.

When a first choice point with an odd number of alternatives (say
$2n+1$) appears, the worker that must own the first alternative (say
$Q$) is given $n+1$ alternatives and the other (say $P$) is given
$n$. The workers then alternate and, in the next choice point, $P$
starts the partitioning. When more choice points with an odd number of
alternatives appear, the split process is repeated. At the end, $Q$
and $P$ may have the same number of unexplored alternatives or, in the
worst case, $Q$ may have one more alternative than $P$. The
pseudo-code for this procedure is shown next in
Fig.~\ref{code_diagonal_splitting}.

\begin{figure}[ht]
\verbatimproperties
\begin{verbatim}
// the sharing worker P starts the adjustment
if (sharing worker) adjust = TRUE else adjust = FALSE
current_cp = top_cp
while(current_cp != root_cp)  // loop until the root choice point
  alt = CP_alt(current_cp)
  if (alt != NULL)
    offset = CP_offset(current_cp)   
    CP_offset(current_cp) = offset * 2
    if (adjust)
      CP_alt(current_cp) = get_next_alternative(alt, offset)
    n_alts = number_of_unexplored_alternatives(alt) / offset
    if (n_alts mod 2 != 0)  // workers alternate
      adjust = !adjust
  current_cp = CP_b(current_cp)  // next choice point on stack
\end{verbatim}
\caption{Work sharing with diagonal splitting}
\label{code_diagonal_splitting}
\end{figure}


\subsection{Incremental Copy}

In YapOr's original implementation, the incremental copy process
copies everything in \emph{P}'s stacks that is missing in
\emph{Q}. With stack splitting, it only copies the segments between
\emph{Q}'s \texttt{top\_cp} before and after sharing for the global
and local stacks. For the trail stack, the copy is the same since this
is necessary to correctly implement the \emph{installation
  phase}~\cite{Rocha-99b}, where $Q$ installs from $P$ the bindings
made to variables belonging to the common segments not copied from
$P$.

\begin{wrapfigure}{r}{8cm}
\vspace{-\intextsep}
\centering
\includegraphics[width=8cm]{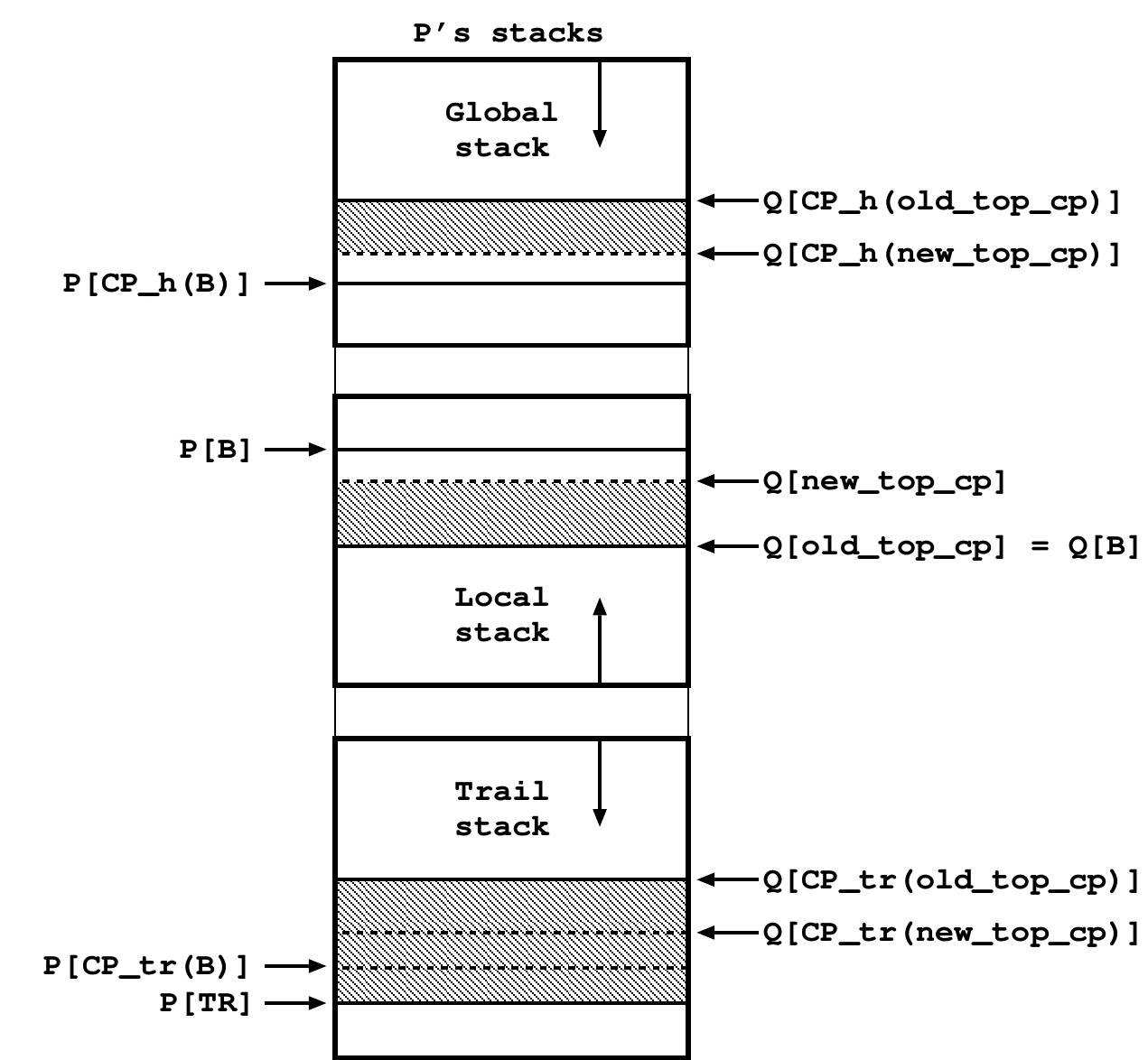}
\caption{Segments to copy with incremental copy}
\label{fig_segments_to_copy}
\vspace{-\intextsep}
\end{wrapfigure}

Figure~\ref{fig_segments_to_copy} illustrates the stack segments to be
copied with incremental copy. For vertical splitting, if \emph{P} has
private work, $Q$'s \texttt{new\_top\_cp} is assigned with the second
choice point in \emph{P}'s choice point set (\texttt{P[CP\_b(B)]}). If
there is no private work, the \texttt{new\_top\_cp} is assigned with
the choice point corresponding to the or-frame pointed by
\texttt{P[OrFr\_nearest\_livenode(CP\_or\_fr(old\_top\_cp))]}. For
half splitting, the \texttt{new\_top\_cp} is always assigned with the
choice point denoted by \texttt{P[CP\_b(middle\_cp)]}. For the
horizontal and diagonal splitting, the assigning ranges are similar to
YapOr's original implementation.

We next discuss the situations where \emph{Q}'s new
\texttt{top\_or\_frame}, assigned during sharing, is older than
\emph{Q}'s \texttt{top\_or\_frame} before sharing. In such case,
\emph{Q} does not copy any segment from \emph{P} and only needs to
move up in the search tree in order to be consistent with the new
assigned \texttt{top\_or\_frame}. In this movement, we may have to
update the or-frames corresponding to the backtracked path by removing
\emph{Q} from the membership fields and by executing a \emph{checking
  phase}. The checking phase is necessary to avoid incoherent values
in the \texttt{CP\_alt} fields in \emph{Q}'s choice points not copied
from \emph{P}. For half splitting, it also avoids incoherent values in
the split counter fields for \emph{Q}'s choice points not copied from
\emph{P}. We can say that such incoherency can be caused by the
independent work sharing operations with different workers that make
the common (not copied) stack segments of \emph{P} and \emph{Q}, to be
inconsistent in \emph{Q}.


\section{Experimental Results}

In this section, we evaluate and compare the performance of the five
splitting strategies on a set of well-known benchmarks. The
environment for our experiments was a multicore machine with 4 AMD
Six-Core Opteron TM 8425 HE (2100 MHz) chips (24 cores in total) and
64 GB of DDR-2 667MHz RAM, running Linux (kernel 2.6.31.5-127 64 bits)
with Yap Prolog 6.3.2. The machine was running in multi-user mode, but
no other users were using it. For the benchmarks, we used the
following set of programs:

\begin{description}
\item [cubes(N)] a program that consists of stacking N colored
  cubes in a column in such a way that no color appears twice in the
  same column for each side.
\item [ham(N)] a program for finding all the Hamiltonian cycles in a
  graph with N nodes, with each node connected to 3 other nodes.
\item [magic(N)] a program to solve the Rubik's magic cube problem in
  N steps.
\item [maze(N)] a program that solves a maze problem in N steps by
  moving an empty square in a 4x4 grid.
\item [nsort(N)] a program for ordering a list of N elements
  using a naive algorithm and starting with the list inverted.
\item [queens(N)] a program to solve the N-queens problem that
  analyzes the board state at every step.
\item [puzzle] a program that solves a puzzle problem where the
  diagonals must add up to the same amount.
\end{description}

All benchmarks find all the solutions for the given problem by
simulating an automatic failure whenever a new solution is found. Each
benchmark was executed 10 consecutive times and the results are the
average of those executions.

We start by measuring the cost of the parallel strategies over the
sequential system. Table~\ref{tab_sequential} presents the execution
times, in seconds, for the set of benchmark programs, when using the
sequential version of Yap and the respective ratios when using the
several parallel models with one worker. In general, for all models,
YapOr overheads result from handling the work load register and from
operations that (i) verify whether the youngest node is shared or
private, (ii) check for sharing requests, and (iii) check for
backtracking messages due to cut operations.

\begin{table}[ht]
\centering
\caption{Execution times, in seconds, for Yap's sequential model and
  the respective overhead ratios for YapOr running 1 worker with
  YapOr's original splitting strategy (OS), vertical splitting (VS),
  half splitting (\nicefrac{1}{2}S), horizontal splitting (HS) and
  diagonal splitting (DS).}
\begin{tabular}{lr|rrrrr}
\hline
\multicolumn{1}{l}{\multirow{4}{*}{\bf Programs}}
& \multicolumn{1}{c}{\multirow{4}{*}{\bf Yap}}
& \multicolumn{5}{c}{\multirow{2}{*}{\bf YapOr~/~Yap}} \\ \\ \cline{3-7}
& \multicolumn{1}{c}{}
& \multicolumn{1}{c}{\multirow{2}{*}{\bf OS}}
& \multicolumn{1}{c}{\multirow{2}{*}{\bf VS}}
& \multicolumn{1}{c}{\multirow{2}{*}{\bf \nicefrac{1}{2}S}}
& \multicolumn{1}{c}{\multirow{2}{*}{\bf HS}}
& \multicolumn{1}{c}{\multirow{2}{*}{\bf DS}} \\ \\
\hline
\bf cubes(7)   & \bw{  0.200} & \bw{1.050} & \bw{1.080} & \bw{1.070} & \bw{1.110} & \bw{1.135}\\
\bf ham(26)    & \bw{  0.350} & \bw{1.169} & \bw{1.180} & \bw{1.177} & \bw{1.094} & \bw{1.100}\\
\bf magic(6)   & \bw{  5.102} & \bw{1.045} & \bw{1.036} & \bw{1.005} & \bw{1.245} & \bw{1.252}\\
\bf magic(7)   & \bw{ 45.865} & \bw{1.051} & \bw{1.021} & \bw{1.007} & \bw{1.251} & \bw{1.261}\\
\bf maze(10)   & \bw{  0.623} & \bw{1.064} & \bw{1.050} & \bw{1.050} & \bw{1.273} & \bw{1.207}\\
\bf maze(12)   & \bw{ 10.558} & \bw{1.057} & \bw{1.041} & \bw{1.035} & \bw{1.268} & \bw{1.214}\\
\bf nsort(10)  & \bw{  2.775} & \bw{1.124} & \bw{1.155} & \bw{1.096} & \bw{1.074} & \bw{1.072}\\
\bf nsort(12)  & \bw{368.862} & \bw{1.128} & \bw{1.074} & \bw{1.057} & \bw{1.081} & \bw{1.082}\\
\bf queens(11) & \bw{  1.216} & \bw{1.039} & \bw{1.234} & \bw{1.051} & \bw{1.036} & \bw{1.107}\\
\bf queens(13) & \bw{ 47.187} & \bw{1.025} & \bw{1.165} & \bw{1.053} & \bw{1.043} & \bw{1.039}\\
\bf puzzle     & \bw{  0.153} & \bw{1.157} & \bw{1.235} & \bw{1.144} & \bw{1.176} & \bw{1.157}\\
\hline
{\bf Average}  &              & \bw{1.083} & \bw{1.116} & \bw{1.068} & \bw{1.150} & \bw{1.148}\\
\hline
\end{tabular}
\label{tab_sequential}
\end{table}

\begin{table}[ht]
\centering
\caption{Speedups for YapOr running 16 and 24 workers with YapOr's
  original splitting strategy (OS), vertical splitting (VS), half
  splitting (\nicefrac{1}{2}S), horizontal splitting (HS) and diagonal
  splitting (DS) without the incremental copy technique.}
\begin{tabular}{lrrrrr|rrrrr}
\hline
\multirow{4}{*}{\bf Programs}
& \multicolumn{5}{c}{\multirow{2}{*}{\bf 16 Workers}} 
& \multicolumn{5}{c}{\multirow{2}{*}{\bf 24 Workers}} \\ \\
\cline{2-6}\cline{7-11}
& \multicolumn{1}{c}{\multirow{2}{*}{\bf OS}} 
& \multicolumn{1}{c}{\multirow{2}{*}{\bf VS}} 
& \multicolumn{1}{c}{\multirow{2}{*}{\bf \nicefrac{1}{2}S}} 
& \multicolumn{1}{c}{\multirow{2}{*}{\bf HS}} 
& \multicolumn{1}{c}{\multirow{2}{*}{\bf DS}}
& \multicolumn{1}{c}{\multirow{2}{*}{\bf OS}} 
& \multicolumn{1}{c}{\multirow{2}{*}{\bf VS}} 
& \multicolumn{1}{c}{\multirow{2}{*}{\bf \nicefrac{1}{2}S}} 
& \multicolumn{1}{c}{\multirow{2}{*}{\bf HS}} 
& \multicolumn{1}{c}{\multirow{2}{*}{\bf DS}} \\ \\
\hline
\bf cubes(7)   & \bg{6.45}  & \bw{4.65}  & \bw{0.61}  & \bw{5.26}  & \bw{5.12}
               & \bg{6.66}  & \bw{3.92}  & \bw{0.46}  & \bw{4.76}  & \bw{4.54}\\
\bf ham(26)    & \bg{6.14}  & \bw{4.86}  & \bw{2.34}  & \bw{4.11}  & \bw{5.14}
               & \bg{6.36}  & \bw{4.79}  & \bw{2.07}  & \bw{3.97}  & \bw{5.14}\\
\bf magic(6)   & \bg{14.33} & \bw{14.25} & \bw{8.35}  & \bw{11.67} & \bw{11.70}
               & \bg{20.40} & \bw{19.77} & \bw{7.76}  & \bw{16.51} & \bw{16.35}\\
\bf magic(7)   & \bw{14.97} & \bg{15.51} & \bw{12.18} & \bw{12.29} & \bw{12.31}
               & \bw{22.24} & \bg{22.96} & \bw{16.17} & \bw{18.39} & \bw{18.43}\\
\bf maze(10)   & \bw{9.58}  & \bg{10.74} & \bw{4.82}  & \bw{7.78}  & \bw{7.98}
               & \bw{11.32} & \bg{11.98} & \bw{4.20}  & \bw{9.16}  & \bw{8.41}\\
\bf maze(12)   & \bw{14.44} & \bg{15.06} & \bw{11.55} & \bw{12.50} & \bw{12.56}
               & \bw{21.03} & \bg{21.81} & \bw{14.89} & \bw{17.80} & \bw{17.68}\\
\bf nsort(10)  & \bw{10.63} & \bg{11.37} & \bw{9.91}  & \bw{9.94}  & \bw{10.16}
               & \bg{13.73} & \bw{12.50} & \bw{12.06} & \bw{12.50} & \bw{12.33}\\
\bf nsort(12)  & \bw{14.37} & \bw{14.71} & \bg{14.72} & \bw{14.43} & \bw{14.52}
               & \bw{21.16} & \bw{21.47} & \bg{21.62} & \bw{20.93} & \bw{20.78}\\
\bf queens(11) & \bg{12.66} & \bw{7.84}  & \bw{1.68}  & \bw{11.05} & \bw{11.15}
               & \bg{16.21} & \bw{8.94}  & \bw{1.60}  & \bw{13.07} & \bw{12.93}\\
\bf queens(13) & \bg{15.66} & \bw{14.05} & \bw{4.10}  & \bw{15.08} & \bw{15.16}
               & \bw{22.14} & \bw{20.54} & \bw{4.12}  & \bw{22.20} & \bg{22.42}\\
\bf puzzle     & \bg{3.82}  & \bw{2.21}  & \bw{2.25}  & \bw{3.00}  & \bw{3.12}
               & \bg{3.73}  & \bw{1.91}  & \bw{1.45}  & \bw{2.59}  & \bw{2.68}\\
\hline
\bf Average    & \bg{11.19} & \bw{10.48} & \bw{6.59}  & \bw{9.74}  & \bw{9.90}
               & \bg{15.00} & \bw{13.69} & \bw{7.85}  & \bw{12.90} & \bw{12.88}\\
\hline
\end{tabular}
\label{tab_speedups_without_ic}
\end{table}

Results in Table~\ref{tab_sequential} show that for these set of
benchmarks, YapOr's overhead with each of the splitting strategies is
small, between 6.8\% and 15\%. This is in-line with the overheads
observed previously for YapOr and some of the splitting
strategies~\cite{Rocha-99b,CostaVS-10,Vieira-12}.

Next, we assessed the performance of the or-parallel models, by
running YapOr with a varying number of workers, up to 24, although for
simplicity here we only show results for 16 and 24 workers. For
fairness in the comparison of all strategies, we use the sequential
execution times as the base execution times, instead of considering
the base execution times with 1 worker for each strategy. In this way,
the speedups do reflect real gains from sequential execution
times. The results are shown in Tables~\ref{tab_speedups_without_ic}
and \ref{tab_speedups_with_ic} and the best speedup value among all
strategies, which corresponds to the fastest execution times, for each
benchmark, is marked with a gray background color.

From Table~\ref{tab_speedups_without_ic} we can observe the overall
performance of all strategies without resorting to incremental copy
optimization. The results show reasonably good speedups with exception
for half splitting. With 24 workers, YapOr's original splitting shows
the best performance, followed by vertical splitting and then
horizontal and diagonal splitting with minimal differences. For some
benchmarks, such as the \textbf{cubes} and \textbf{queens} benchmarks,
half splitting does pretty badly.

Table~\ref{tab_speedups_with_ic} shows the overall performance for all
strategies, but now using the incremental copying optimization. The
performance for all strategies improve significantly for all
benchmarks. Again, half splitting is the worst performing strategy, on
average, it performs about 14\% less than the best performing strategy
with 24 workers. Another observation is that vertical, horizontal and
diagonal splitting perform slightly close to the original YapOr. The
best overall performance with 16 and 24 workers is achieved with
vertical splitting.

\begin{table}[ht]
\centering
\caption{Speedups for YapOr running 16 and 24 workers with YapOr's
  original splitting strategy (OS), vertical splitting (VS), half
  splitting (\nicefrac{1}{2}S), horizontal splitting (HS) and
  diagonal splitting (DS) with the incremental copy technique.}
\begin{tabular}{lrrrrr|rrrrr}
\hline
\multirow{4}{*}{\bf Programs}
& \multicolumn{5}{c}{\multirow{2}{*}{\bf 16 Workers}} 
& \multicolumn{5}{c}{\multirow{2}{*}{\bf 24 Workers}} \\ \\ \cline{2-6}\cline{7-11}
& \multicolumn{1}{c}{\multirow{2}{*}{\bf OS}} 
& \multicolumn{1}{c}{\multirow{2}{*}{\bf VS}} 
& \multicolumn{1}{c}{\multirow{2}{*}{\bf \nicefrac{1}{2}S}} 
& \multicolumn{1}{c}{\multirow{2}{*}{\bf HS}} 
& \multicolumn{1}{c}{\multirow{2}{*}{\bf DS}}
& \multicolumn{1}{c}{\multirow{2}{*}{\bf OS}} 
& \multicolumn{1}{c}{\multirow{2}{*}{\bf VS}} 
& \multicolumn{1}{c}{\multirow{2}{*}{\bf \nicefrac{1}{2}S}} 
& \multicolumn{1}{c}{\multirow{2}{*}{\bf HS}} 
& \multicolumn{1}{c}{\multirow{2}{*}{\bf DS}} \\ \\
\hline
\bf cubes(7)   & \bw{8.00}  & \bg{13.33} & \bw{6.45}  & \bg{13.33} & \bw{12.50}
               & \bw{13.33} & \bw{14.28} & \bw{4.00}  & \bg{16.66} & \bw{15.38}\\
\bf ham(26)    & \bw{10.00} & \bw{10.29} & \bw{7.95}  & \bw{10.00} & \bg{11.29}
               & \bg{9.45}  & \bw{7.60}  & \bw{4.48}  & \bw{7.14}  & \bg{9.45}\\
\bf magic(6)   & \bw{14.96} & \bg{15.46} & \bw{15.27} & \bw{12.41} & \bw{12.47}
               & \bw{22.08} & \bg{22.87} & \bw{22.77} & \bw{18.41} & \bw{18.41}\\
\bf magic(7)   & \bw{15.15} & \bg{15.64} & \bw{15.46} & \bw{12.52} & \bw{12.50}
               & \bw{22.63} & \bg{23.40} & \bw{22.96} & \bw{18.67} & \bw{18.78}\\
\bf maze(10)   & \bw{13.54} & \bg{15.19} & \bw{14.83} & \bw{12.46} & \bw{12.71}
               & \bw{18.32} & \bg{22.25} & \bw{21.48} & \bw{18.32} & \bw{18.87}\\
\bf maze(12)   & \bw{15.12} & \bg{15.59} & \bw{15.25} & \bw{13.18} & \bw{13.46}
               & \bw{22.36} & \bg{23.30} & \bw{22.75} & \bw{19.73} & \bw{19.95}\\
\bf nsort(10)  & \bw{14.15} & \bg{14.60} & \bg{14.60} & \bw{14.15} & \bw{14.08}
               & \bw{20.25} & \bw{20.70} & \bg{21.34} & \bw{19.96} & \bw{20.40}\\
\bf nsort(12)  & \bw{14.18} & \bw{14.36} & \bg{14.43} & \bw{14.04} & \bw{14.26}
               & \bw{21.59} & \bg{22.28} & \bw{22.16} & \bw{21.69} & \bw{21.85}\\
\bf queens(11) & \bw{14.65} & \bw{13.66} & \bw{9.57}  & \bg{14.82} & \bg{14.82}
               & \bw{20.26} & \bw{17.62} & \bw{6.75}  & \bw{20.26} & \bg{20.96}\\
\bf queens(13) & \bg{15.75} & \bw{14.51} & \bw{13.87} & \bw{15.35} & \bw{15.32}
               & \bg{23.44} & \bw{21.60} & \bw{15.90} & \bw{22.99} & \bw{22.91}\\
\bf puzzle     & \bw{9.00}  & \bw{10.20} & \bg{11.76} & \bg{11.76} & \bg{11.76}
               & \bw{9.56}  & \bw{10.20} & \bg{15.30} & \bw{10.92} & \bw{12.75}\\
\hline
\bf Average    & \bw{13.13} & \bg{13.89} & \bw{12.68} & \bw{13.09} & \bw{13.20}
               & \bw{18.48} & \bg{18.74} & \bw{16.35} & \bw{17.71} & \bw{18.16}\\
\hline
\end{tabular}
\label{tab_speedups_with_ic}
\end{table}

Instead of using the sequential execution times as the base reference,
if one uses the execution times with 1 worker for each strategy, then
the average speedups with incremental copying and 24 workers for the
original, vertical, horizontal and diagonal splitting were very close
and above 20.


\section{Conclusions and Further Work}

We have presented the integration of five alternative splitting
strategies on top of the YapOr system for or-parallel Prolog execution
on multicores. Our implementation shares the underlying execution
environment and most of the data structures used to implement
or-parallelism in YapOr.

Experimental results, on a multicore machine with 24 cores, showed
that clearly incremental copying optimization pays off in improving
real performance in all strategies. The results for all strategies are
reasonably good and the average speedups over all benchmarks is
reasonably close, with exception for half splitting that performs a
little worse. However, these are preliminary results and further
detailed statistics are necessary to enable us to explain some
apparently inconsistent results. For example, half splitting performs
badly with \textbf{cubes} and \textbf{queens} benchmarks, both with
incremental and without incremental copying, but, on the other hand,
it is the best performing on the \textbf{nsort(10)} and
\textbf{puzzle} benchmarks with incremental copying. To explain these
results, we need not only to gather low level statistics, during the
execution, but also understand in which manner the splitting strategy
influences the scheduling of work. A postmortem visualization of the
search tree might also bring some insight in to this analysis.

Although stack splitting was initially proposed for distributed memory
architectures, the results show that it is equally suitable for
multicore architectures. This is an interesting advantage of stack
splitting since we could use it as the basis for a hybrid execution
model aiming at clusters of multicores. The idea is to combine workers
into teams. A team of workers might run on shared memory and use any
splitting strategy to distribute work. Different teams might be
assigned to different cluster nodes and can distribute work using
stack splitting.


\section*{Acknowledgments}

We thank the referees for their valuable comments and
suggestions. This work is partially funded by the ERDF (European
Regional Development Fund) through the COMPETE Programme and by FCT
(Portuguese Foundation for Science and Technology) within projects
PEst (FCOMP-01-0124-FEDER-022701), LEAP (PTDC/EIA-CCO/112158/2009) and
HORUS (PTDC/EIA-EIA/100897/2008).


\bibliographystyle{splncs}
\bibliography{references}

\begin{thebibliography}{10}

\bibitem{Ali-90a}
Ali, K., Karlsson, R.:
\newblock {The Muse Approach to OR-Parallel Prolog}.
\newblock International Journal of Parallel Programming \textbf{19}(2) (1990)
  129--162

\bibitem{Rocha-99b}
Rocha, R., Silva, F., {Santos Costa}, V.:
\newblock {YapOr: an Or-Parallel Prolog System Based on Environment Copying}.
\newblock In: Portuguese Conference on Artificial Intelligence. Number 1695 in
  LNAI, Springer-Verlag (1999)  178--192

\bibitem{Ali-90b}
Ali, K., Karlsson, R.:
\newblock {Full Prolog and Scheduling OR-Parallelism in Muse}.
\newblock International Journal of Parallel Programming \textbf{19}(6) (1990)
  445--475

\bibitem{Gupta-99}
Gupta, G., Pontelli, E.:
\newblock {Stack Splitting: A Simple Technique for Implementing Or-parallelism
  on Distributed Machines}.
\newblock In: International Conference on Logic Programming, The MIT Press
  (1999)  290--304

\bibitem{Pontelli-06}
Pontelli, E., Villaverde, K., Guo, H.F., Gupta, G.:
\newblock {Stack splitting: A technique for efficient exploitation of search
  parallelism on share-nothing platforms}.
\newblock Journal of Parallel and Distributed Computing \textbf{66}(10) (2006)
  1267--1293

\bibitem{Villaverde-01}
Villaverde, K., Pontelli, E., Guo, H., Gupta, G.:
\newblock {PALS: An Or-Parallel Implementation of Prolog on Beowulf
  Architectures}.
\newblock In: International Conference on Logic Programming. Number 2237 in
  LNCS, Springer-Verlag (2001)  27--42

\bibitem{Rocha-03a}
Rocha, R., Silva, F., Martins, R.:
\newblock {YapDss: an Or-Parallel Prolog System for Scalable Beowulf Clusters}.
\newblock In: Portuguese Conference on Artificial Intelligence. Number 2902 in
  LNAI, Springer-Verlag (2003)  136--150

\bibitem{Vieira-12}
Vieira, R., Rocha, R., Silva, F.:
\newblock {Or-Parallel Prolog Execution on Multicores Based on Stack
  Splitting}.
\newblock In: International Workshop on Declarative Aspects and Applications of
  Multicore Programming, ACM Digital Library (2012)

\bibitem{CostaVS-12}
{Santos Costa}, V., Rocha, R., Damas, L.:
\newblock {The YAP Prolog System}.
\newblock Journal of Theory and Practice of Logic Programming \textbf{12}(1 \&
  2) (2012)  5--34

\bibitem{Villaverde-03}
Villaverde, K., Pontelli, E., Guo, H., Gupta, G.:
\newblock {A Methodology for Order-Sensitive Execution of Non-deterministic
  Languages on Beowulf Platforms}.
\newblock In: International Euro-Par Conference. Number 2790 in LNCS,
  Springer-Verlag (2003)  694--703

\bibitem{CostaVS-10}
{Santos Costa}, V., Dutra, I., Rocha, R.:
\newblock {Threads and Or-Parallelism Unified}.
\newblock Journal of Theory and Practice of Logic Programming, International
  Conference on Logic Programming, Special Issue \textbf{10}(4--6) (2010)
  417--432

\end{thebibliography}


\end{document}